\documentclass[twocolumn]{aastex62}
\usepackage{epsfig}
\usepackage{natbib}
\usepackage{amsmath}
\usepackage{color}
\usepackage{graphicx}

\newcommand{\kms}{km~s$^{-1}$}

\newcommand{\kmsMpc}{km~s$^{-1}$~Mpc$^{-1}$}
\DeclareMathAlphabet{\mathpzc}{OT1}{pzc}{m}{it}

\shorttitle{Peculiar Velocities Just Beyond the Local Group}
\shortauthors{Anand et al.}

\begin{document}

\title{Peculiar Velocities of Galaxies Just Beyond the Local Group}

\author{Gagandeep S. Anand}
\affiliation{Institute for Astronomy, University of Hawaii, 2680 Woodlawn Drive, Honolulu, HI 96822, USA}

\author{R. Brent Tully}
\affiliation{Institute for Astronomy, University of Hawaii, 2680 Woodlawn Drive, Honolulu, HI 96822, USA}

\author{Luca Rizzi}
\affiliation{W.M. Keck Observatory, 65-1120 Mamalahoa Highway, Kamuela, HI 96743, USA}

\author{Edward J. Shaya}
\affiliation{Astronomy Department, University of Maryland, College Park, MD 20743, USA}

\author{Igor D. Karachentsev}
\affiliation{Special Astrophysical Observatory, Nizhniy Arkhyz, Karachai-Cherkessia 369167, Russia}

\begin{abstract}
The Milky Way lies in a thin plane, the Local Sheet, a part of a wall bounding the Local Void lying toward the north supergalactic pole.  Galaxies with accurate distances both above and below this supergalactic equatorial plane have systematically negative peculiar velocities. The interpretation of this situation is that the Local Void is expanding, giving members of the Local Sheet deviant velocities toward the south supergalactic pole.  The few galaxies within the void are evacuating the void.  Galaxies in a filament in the supergalactic south are not feeling the expansion so their apparent motion toward us is mainly a reflex of our motion. The model of the local velocity field was uncertain because the apex of our motion away from the Local Void lies in obscurity near the Galactic plane.  Here, results of Hubble Space Telescope infrared observations are reported that find tip of the red giant branch distances to four obscured galaxies.  All the distances are $\sim7$~Mpc, revealing that these galaxies are part of a continuous filamentary structure passing between the north and south Galactic hemispheres and sharing the same kinematic signature of peculiar velocities toward us. A fifth galaxy nearby in projection, GALFA-DW4, has an ambiguous distance.  If nearby at $\sim 3$~Mpc, this galaxy has an anomalous velocity {\it away} from us of +348~\kms.  Alternatively, perhaps the resolved stars are on the asymptotic giant branch and the galaxy is beyond 6~Mpc whence the observed velocity would not be unusual.

\bigskip
\end{abstract}

\smallskip

\section{Introduction}

Thanks to imaging with Hubble Space Telescope (HST), there is an increasingly complete inventory of accurate distances to nearby galaxies derived from the luminosities of stars at the tip of the red giant branch (TRGB).   Typical uncertainties in distances are 5\% for systems within 10~Mpc.   A galaxy at 5~Mpc has an expectation velocity of 375~\kms\ assuming a value of the Hubble Constant of 75~\kmsMpc\ consistent with the TRGB zero-point scale \citep{2016AJ....152...50T}.  Deviations from this expectation, so-called peculiar velocities, can be determined with uncertainties of $\sim 20$~\kms.   In these nearby cases the peculiar velocities can be easily extracted with significance from the Hubble flow, unlike with galaxies at substantial distances where peculiar velocity components tend to be small fractions of the observed velocities.

There are two dominant general characteristics of the motions of nearby galaxies.  Most nearby galaxies lie in a thin plane (the Local Sheet) which significantly motivated the specification of the equatorial equator in supergalactic coordinates  \citep{1956VA......2.1584D}.  Dominant characteristic {\it one} is that galaxies within this plane experience only modest velocity excursions from Hubble flow; only $\sim 40$~\kms\ in the line of sight if the virial regions are excluded  \citep{2002A&A...389..812K, 2003A&A...398..479K}.

\begin{table*}[ht]
\begin{tabular}{|l|c|c|c|c|c|c|c|c|}
\hline
\textbf{Galaxy}      & \textbf{RA (J2000)} & \textbf{Dec (J2000)} & \textit{\textbf{l}} (Gal.) & \textit{\textbf{b}} (Gal.) & \textit{\textbf{sgl}} (SGal.) & \textit{\textbf{sgb}} (SGal.) & \textbf{F110W (s)} & \textbf{F160W (s)} \\ \hline
\textbf{ESO 558-011} & 07:06:57.1          & -22:02:25            & 234.3388   & -6.6215    & 182.8833     & -83.0691     & 3824               & 3824               \\ \hline
\textbf{HIZSS-003}   & 07:00:26.5          & -04:12:37            & 217.7050   & 0.0871     & 33.2401      & -78.4203     & 3824               & 3521               \\ \hline
\textbf{HIZSS-021}   & 07:46:16.2          & -28:28:04            & 244.1748   & -1.8423    & 165.7527     & -72.6546     & 3824               & 3824               \\ \hline
\textbf{Orion Dwarf} & 05:45:02.1          & +05:04:11            & 200.6191   & -12.3003   & 345.2606     & -62.9516     & 2611               & 2611               \\ \hline
\end{tabular}
\caption{Summary of the useable HST WFC3/IR observations from proposal 15150 (PI R. Tully). Each target was observed for 1.5 orbits in each filter (3824 seconds), but two targets (HIZSS-003 and the Orion Dwarf Galaxy) have exposures that are not useable due to excessive drift caused by the failure of HST to properly acquire guide stars. The option to repeat these observations was not pursued due to adequate depth in the useable data.}
\label{15150observations}
\end{table*}

Dominant characteristic {\it two} is an outflow from the Local Void \citep{1987nga..book.....T}.   The Local Sheet is a wall of the Local Void and the expansion of the void gives a motion of the wall normal to the plane, toward the supergalactic south pole \citep{2008ApJ...676..184T}.  The general features of this situation are reasonably well established.  Galaxies both above and below the Local Sheet have peculiar velocities toward us.   Below the supergalactic equatorial plane (negative SGZ) there is a wispy filament that runs from the Leo (NGC3368) Group near the Virgo Cluster through to the Fornax Cluster.  This filament is identified in the Nearby Galaxies Atlas \citep{1987nga..book.....T} as the Leo Spur north of the Galactic plane and as the Dorado Cloud south of the Galactic plane.  Galaxies in this structure are systematically moving toward us relative to the Hubble flow at several hundred \kms\ \citep{2015ApJ...805..144K}.  Positive SGZ is the domain of the Local Void, hence is scantily populated.  Two lonely dwarfs are observed to be rushing toward us \citep{2017ApJ...835...78R}.  The overall situation is captured by a numerical action orbit reconstruction in the volume within 2850~\kms\ \citep{2017ApJ...850..207S}.  In this model, the Local Sheet, our Galaxy included,  is descending in SGZ due to the expansion of the Local Void and the few galaxies within the void at positive SGZ are being expelled even faster, catching up to us.  The galaxies in the Leo Spur$-$Dorado Cloud are not affected by the Local Void expansion so their apparent negative values are a recoil of our downward motion.

While the picture just described was becoming compelling, it was unfortunate that the galaxies with good distances at negative SGZ were significantly displaced from the south supergalactic pole.  Vector decompositions provided only partial information on the vertical (SGZ) components of peculiar velocities.  This circumstance arose because the south supergalactic pole lies in obscuration near the Galactic plane.   This paper describes a successful effort to acquire distances to galaxies at low Galactic latitudes near the supergalactic south pole through imaging with HST at near infrared bands.  The overall results are in good agreement with expectations from the \citet{2017ApJ...850..207S} model.  However there is the possibility that one galaxy has a remarkably anomalous velocity.


\section{Data $\&$ Observations}

\subsection{Existing Data}

Our knowledge of nearby galaxy distances and motions has been tremendously enhanced with deep, high-resolution observations enabled by HST, both in the optical and near-infrared. Over the years, the community has proposed and received observations of $\sim$500 galaxies that have the right combination of depth and filters to successfully extract a TRGB distance. However, a uniform sample with minimal competing systematics requires the data to be photometered and analyzed in the same way. To this end, our collaboration has been maintaining the CMDs/TRGB catalog on our Extragalactic Distance Database, or EDD\footnote{\url{edd.ifa.hawaii.edu}} \citep{2009AJ....138..323T,2009AJ....138..332J}. We currently have analyzed over 450 galaxy CMDs, nearly doubling the initial release of 250. More are being added on a continued basis as observations become available from our own programs and within the HST archives.

\subsection{New HST Observations}

\begin{figure*}
\centering
\gridline{\fig{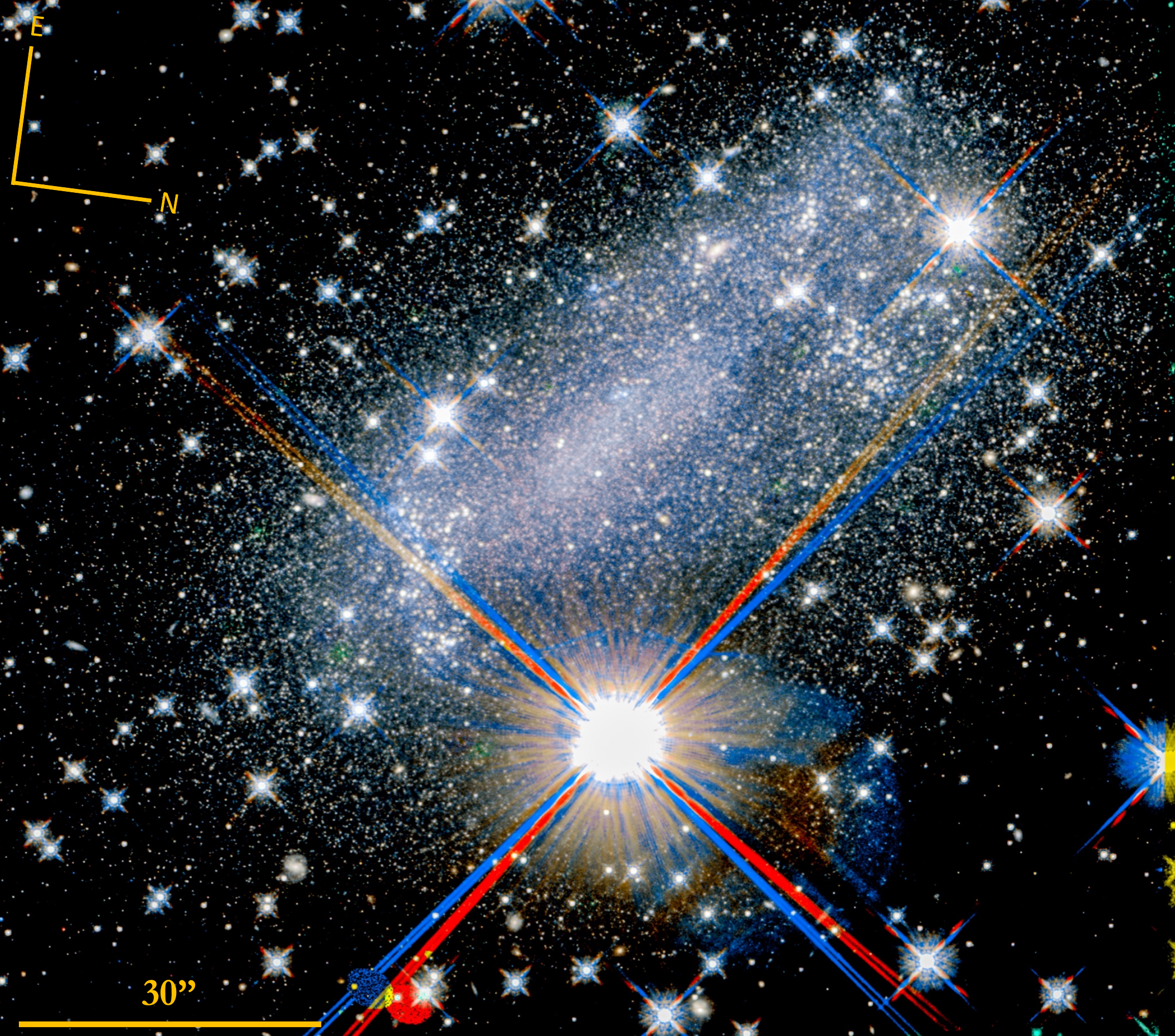}{0.5\textwidth}{(a)}
          \fig{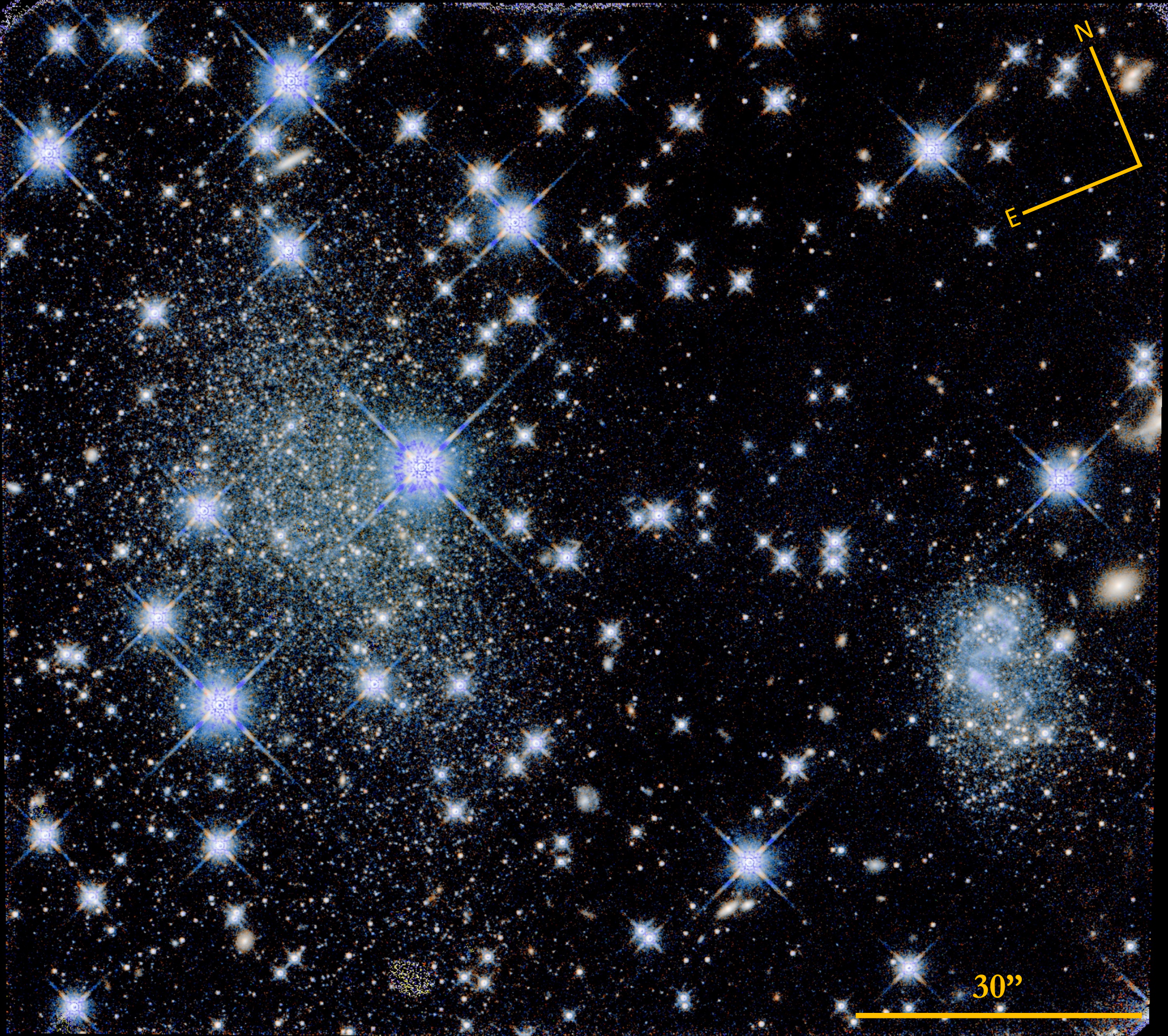}{0.5\textwidth}{(b)}}
\gridline{\fig{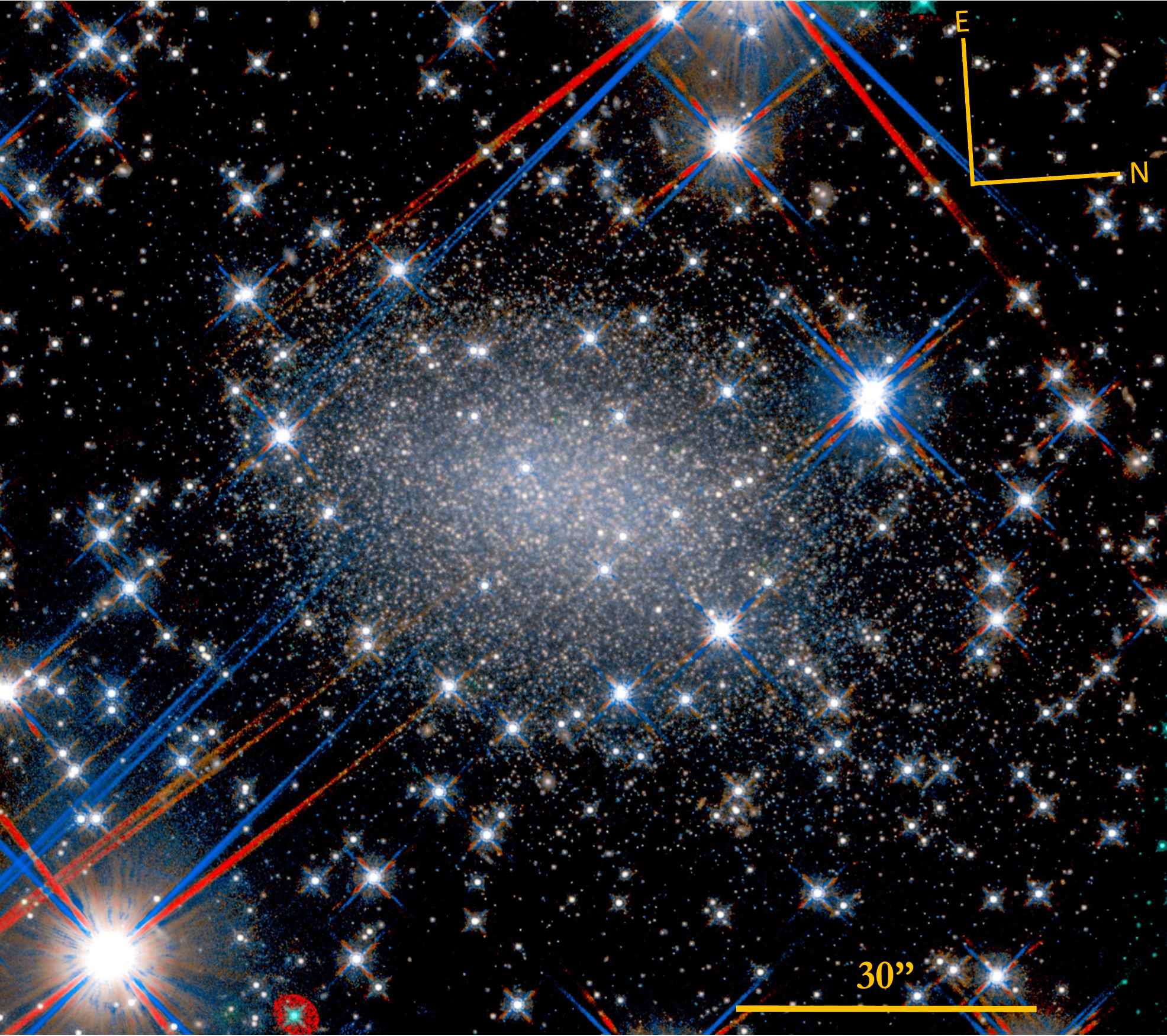}{0.5\textwidth}{(c)}
          \fig{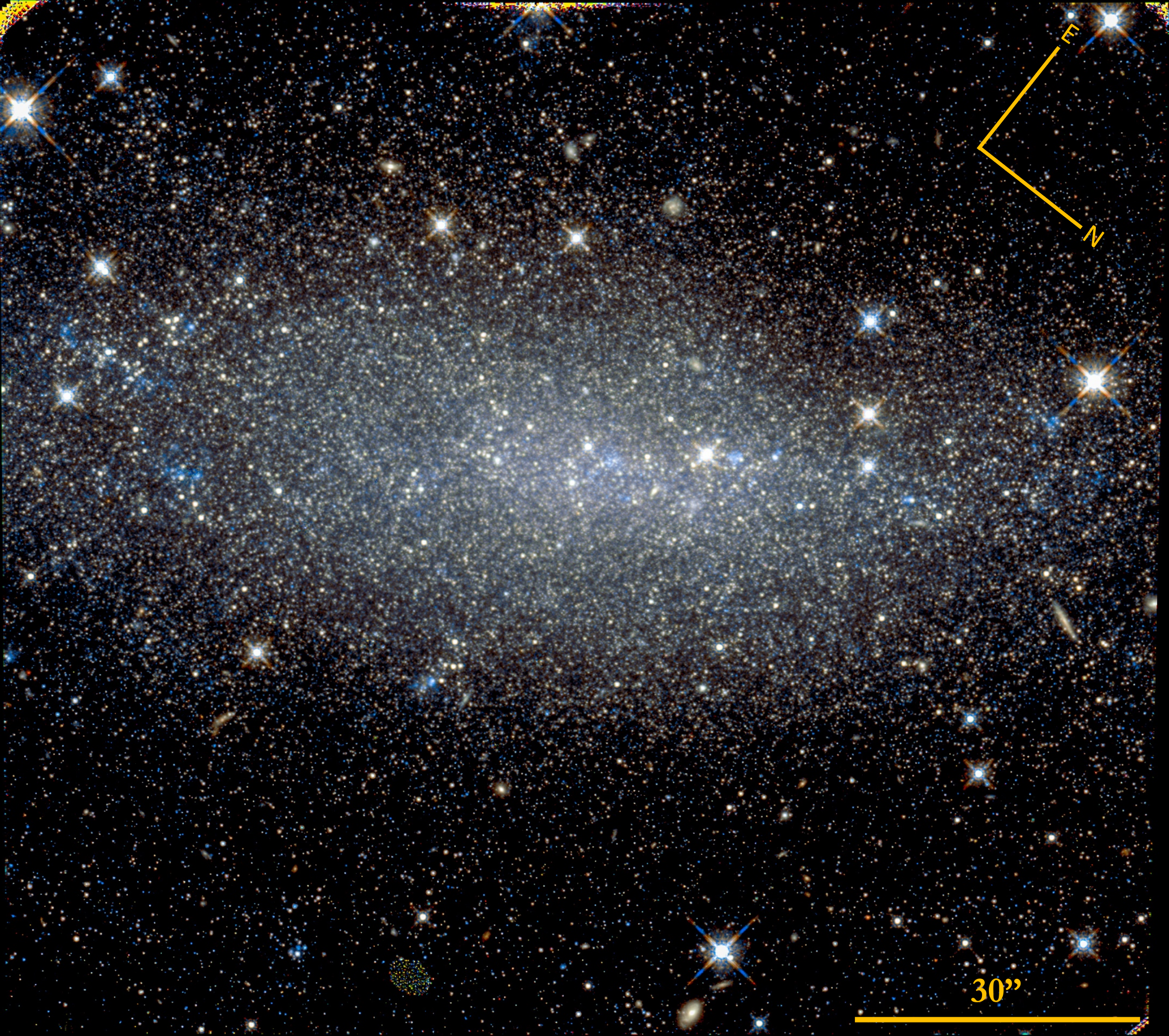}{0.5\textwidth}{(d)}}
\caption{RGB images of the four galaxies (a) ESO 558-011 b) HIZSS-003, c) HIZSS-021, d) Orion Dwarf) constructed from our WFC3/IR F110W+F160W observations. F110W is mapped to blue, F160W is mapped to red, and an average of the two filters is used for green. The offset in colors of the star spikes for some of the targets are due to differing orientations between individual exposures.}
\label{colorImages}
\end{figure*}

The lack of galaxies with accurate distances near the south supergalactic pole prompted us to submit HST proposal 15150 (Cycle 25, PI R. Tully). Four targets were selected, each suspected to lie several megaparsecs beyond their apparent Hubble flow distances ($v/H_{0}$) of 3-4 Mpc. Due to the significant extinction associated with looking through the galactic plane, the galaxies were observed in the near-infrared with WFC3/IR in both F110W $\&$ F160W. 

Each target was observed for 1.5 orbits in each filter, although not all of this was usable. Excessive drift caused by the failure of HST to properly acquire guide stars resulted in a loss of 0.5 orbits in each filter for the Orion Dwarf Galaxy, and 0.25 orbits in F160W for HIZSS-003. The option to repeat these observations was not pursued since the data had reached the necessary depth for our purposes. A summary of the observations is presented in Table \ref{15150observations}. 

Color images for our targets are shown in Figure \ref{colorImages}. We map F160W to red, F110W to blue, and average both filters to get green. The images are aligned and drizzled using DrizzlePac 2.0 \citep{2015ASPC..495..281A}, and combined with APLpy \citep{2012ascl.soft08017R} $\&$ Montage \citep{2010ascl.soft10036J, 2017ASPC..512...81B}, followed by minor adjustments in Lightroom. 


\clearpage

\section{TRGB Measurements}

\subsection{Background}
Stars near the tip of the red giant branch are useful as standard candles due to their relatively uniform luminosities before undergoing the helium flash \citep{1990AJ....100..162D,1993ApJ...417..553L}. The required metallicity corrections are small in the optical bands \citep{2007ApJ...661..815R}, but become more significant in the near-infrared \citep{2012ApJS..198....6D,2014AJ....148....7W}. The method benefits from the fact that there is no requirement of a temporal baseline (such as with Cepheid variables), and that red giant branch (RGB) stars are plentiful in the halos of galaxies, where issues with host galaxy extinction are minimized. For a detailed review of the method, its history, and systematics, see \cite{2018SSRv..214..113B}.

\subsection{Overall Method}
To measure the location of the TRGB, we use a method developed by \cite{2006AJ....132.2729M} and updated by \cite{2014AJ....148....7W}. PSF photometry is performed with DOLPHOT \citep{2000PASP..112.1383D,2016ascl.soft08013D} and culled to include only sources of high quality. Artificial star experiments are also performed within DOLPHOT to quantify the levels of photometric error, bias, and completeness. The luminosity functions of the RGB and asymptotic giant branch (AGB) populations are fitted by a broken power law, with the break signifying the location of the TRGB. The magnitude and color of the TRGB are corrected for galactic extinction, and the appropriate filter-dependent calibrations are applied to produce a final distance modulus to the galaxy.

In general, the procedure for the optical (either ACS/WFC or WFC3/UVIS) and near-infrared (WFC3/IR) reductions and TRGB measurements are very similar. We use separate criteria for culling the photometry in the optical \citep{2017AJ....154...51M} versus the near-infrared \citep{2012ApJS..198....6D}, though we increase the SNR requirement for the F606W optical photometry from 2 to 5. The TRGB is usually only measured in the F814W filter for optical data sets. In the near-infrared, it is useful to measure the TRGB in both F110W and F160W. This is because while F110W suffers more extinction than F160W, the metallicity dependence of the TRGB is nearly half \citep{2014AJ....148....7W}.

The remainder of this section highlights specific modifications/additions to the above well-established procedures, and then presents the resulting distances. The DOLPHOT photometry, CMDs, tables of relevant values, and color images from this work are all available on EDD.

\subsection{Spatial Selections}
One issue that often occurs is confusion of the TRGB by stars with similar magnitudes and colors, including AGB stars and supergiants. These stars are preferentially located in regions with higher host galaxy extinction, and thus may become reddened to the point which they appear on the CMD as stars near the tip of the red giant branch. 

These issues are enhanced when working in the near-infrared, as the RGB and AGB exhibit nearly identical colors. Additionally, the WFC3/IR detector has a relatively large pixel size of 0.13$\arcsec$ (compared to 0.05$\arcsec$ with ACS/WFC), leading to enhanced effects of crowding in the central regions of galaxies. Such effects can lead to significant errors in the final distance \citep{2019ApJ...872L...4A}. To avoid these issues, we use \textit{glue}\footnote{\url{http://glueviz.org/}} to  select out the main body of the galaxy, and remove the associated stars from the galaxy's CMD. The result is a substantial sharpening of the break associated with the TRGB, leading to more significant detections.

\subsection{Extinction}

\begin{figure}
\plotone{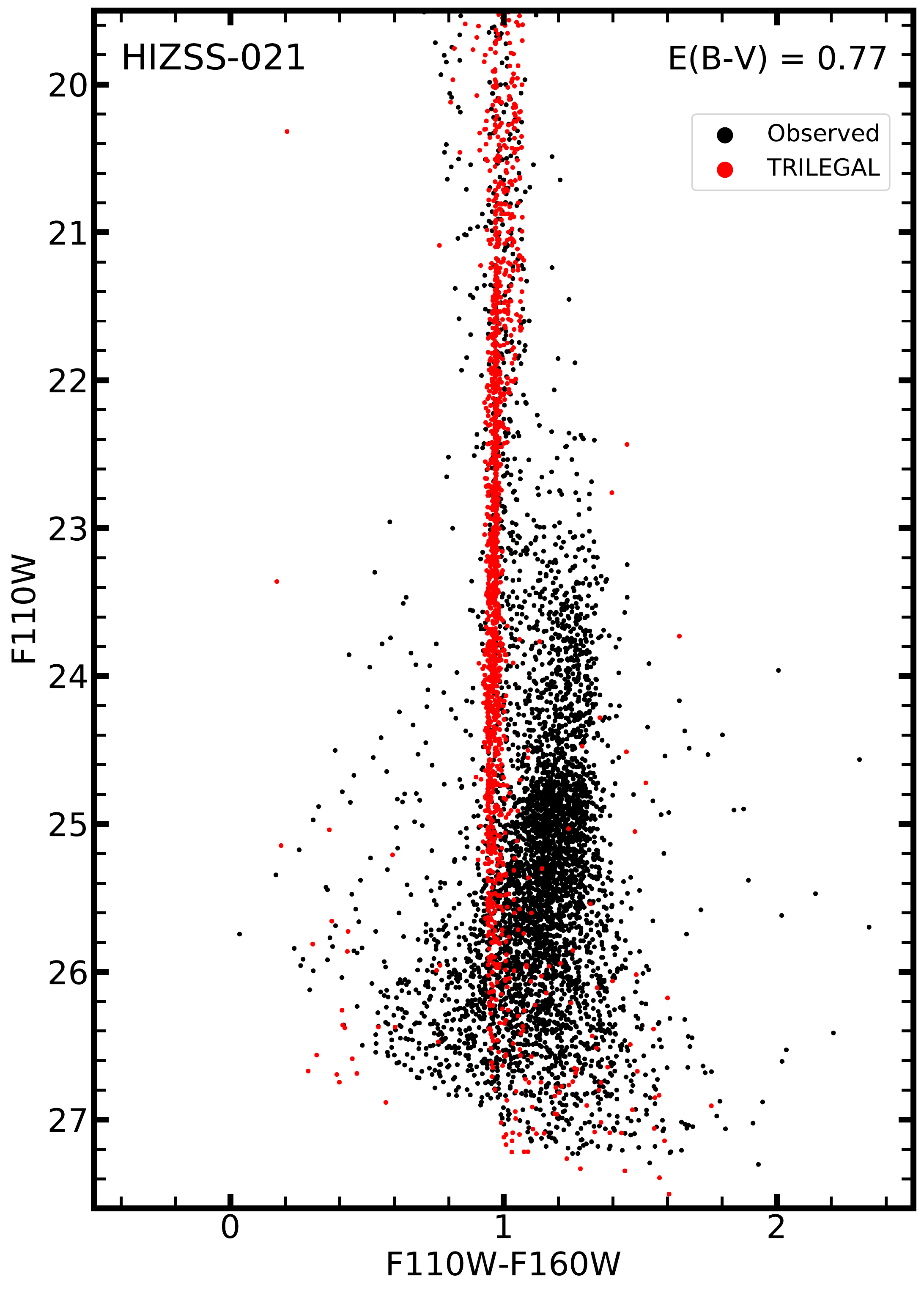}
\caption{Color-magnitude diagram for HIZSS-021 for the full WFC/IR region. Foreground extinction is calculated by simulating non-extincted galactic stars with TRILEGAL, and finding the  extinction value that produces the best-fit to the observed population.}
\label{trilegal}
\end{figure}

By their selection, our four new targets lie at low galactic latitudes. This introduces a possibility that foreground extinction values differ greatly from what is tabulated in all-sky dust maps  \citep{2011ApJ...737..103S}.

To get around this, we use the numerous foreground stars present in the CMDs to our advantage. We employ a technique adapted from \cite{2014AJ....148....7W} -- in brief, we use TRILEGAL \citep{2005A&A...436..895G} to generate a population of unextincted galactic stars in our CMD based on the galactic longitude, latitude, and size of our target fields. The simulated foreground stars are added to the existing CMD, and E(B-V) is varied until a best-fit is achieved to the observed population, with the errors given by the range of possible fits. To maximize the number of observed stars, we use the full field CMDs for this procedure. A sample CMD with the simulated TRILEGAL stars and best-fit is shown in Figure \ref{trilegal}.

In one case (the Orion Dwarf), there are not enough foreground stars to perform this procedure adequately. This is the target most removed from the galactic plane ($b=-12.30^{\circ}$). Fortunately, the Orion Dwarf has a large number of upper main sequence stars detected in its CMD. Following the method of \cite{2014AJ....148....7W}, we use a Sobel filter to calculate the offset of the zero-age main sequence (ZAMS) in color space from NGC 300, a galaxy with a well known E(B-V) = 0.01. We use this offset to find $E(B-V) = 0.595 \pm 0.05$, with the error set to the bin size of the Sobel filter.

\subsection{Distances}
As mentioned above, we obtain measurements of the TRGB magnitude in both F110W and F160W. The magnitudes and colors are corrected for foreground extinction, and the calibrations of \cite{2014AJ....148....7W} are applied to determine the absolute magnitude of the TRGB in each filter.

\begin{figure*}
\plotone{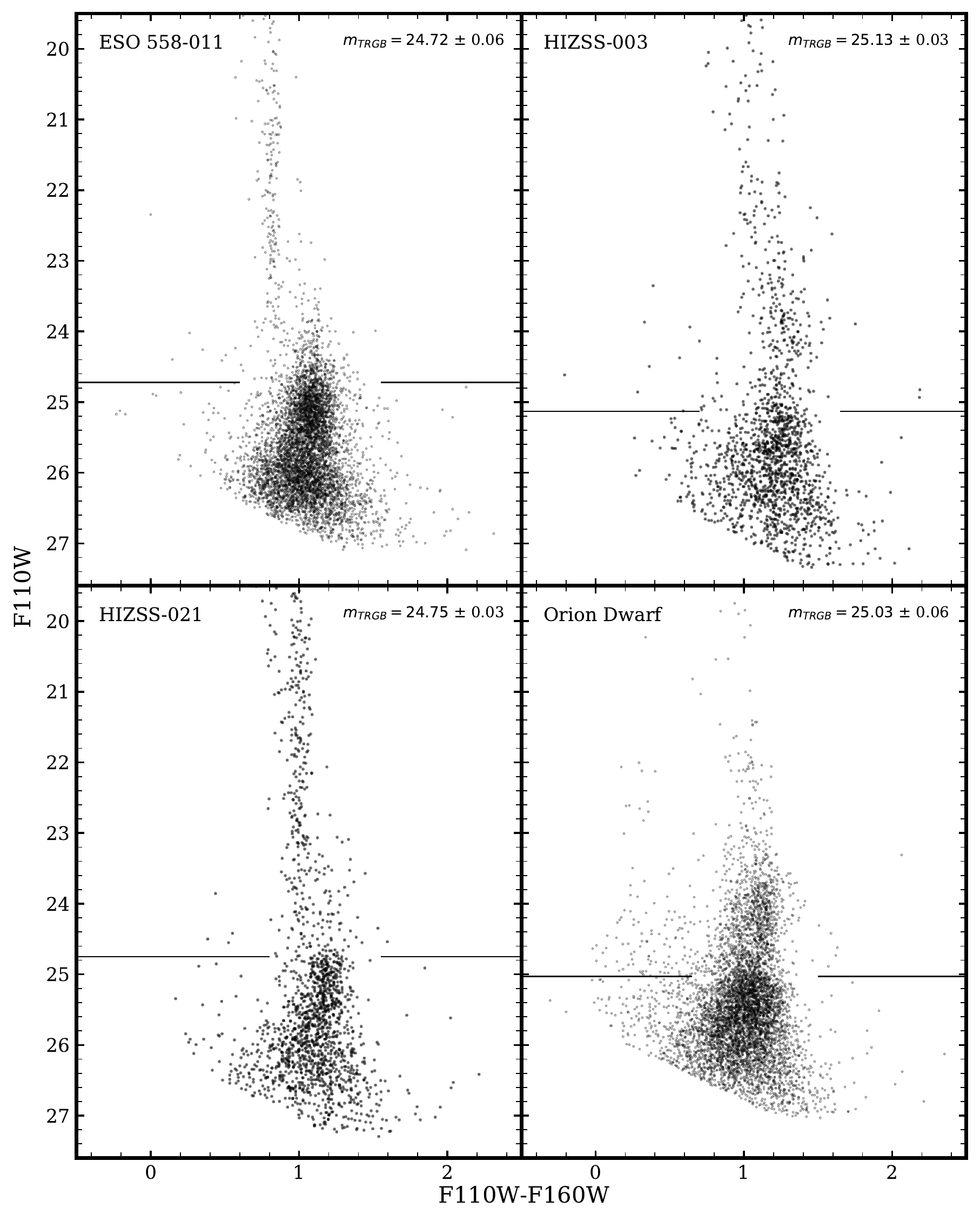}
\caption{F110W vs. F110W-F160W color magnitude diagrams used for finding distances to our four new targets. The TRGB magnitude is shown with the horizontal line, with the break signifying the color range used for the fit.}
\label{4panelTRGB}
\end{figure*}

The spatially selected F110W vs. F110W-F160W CMDs, along with the TRGB measurements in the F110W filter are shown in Figure \ref{4panelTRGB}. A table of measured extinction values, final distances, and derived velocity information is given in Table \ref{distances}. The full-field CMDs, spatially clipped CMDs (including those in F160W), and other relevant measurements are available on EDD. In general, there is good agreement between distances calculated from either magnitude (F110W or F160W), with the maximum discrepancies at the $\sim$2$\%$ level. The agreement is expected since the photometric input is the same and disagreements should only arise from noise. \newline

\subsection{Uncertainty Estimates}
Typically, uncertainties are formally carried through using standard error propagation techniques such as adding in quadrature. For our four near-infrared targets, we decide that this method is likely an underestimate. In the optical bands (e.g. F814W), the absolute magnitude of the TRGB becomes fainter with a redder color. If the estimated foreground extinction is incorrect, the effects on $m_{TRGB}$ and $M_{TRGB}$  act in the opposite directions, so that the error in distance is low, especially given the relatively weak metallicity/color dependence in F814W.

In the near-infrared (e.g. F110W and F160W), $M_{TRGB}$ becomes brighter with redder colors, so that an error in the foreground extinction acts in the same direction upon $m_{TRGB}$ and $M_{TRGB}$. Combining this with the stronger color dependence in the near-infrared, the result is more significant errors in the final distance. Given this, we opt to use the full range of measured TRGB magnitudes, colors, and E(B-V) values in our analysis. The final distance error estimates in Table \ref{distances} reflect the range of determined values. 

\subsection{Previous Literature}
Two of our dwarf galaxies (HIZSS-003 and Orion) have been the topic of modest discussion in the literature. Our new and accurate distances provide valuable insight into these Local Volume dwarfs.

\subsubsection{HIZSS-003}
\cite{2003AJ....126.2362M} discovered an optical (H$\alpha$) counterpart to the HI cloud HIZSS-003 \citep{2000AJ....119.2686H}. A tentative distance was later provided by \citep{2005ApJ...623..148S}, who present near-infrared VLT observations (J+K) and use them to derive a TRGB distance. They report d = 1.69 $\pm$ 0.07 Mpc, which is almost 4$\times$ closer than the distance we present. Their Figure 6 shows the very sparse CMD used for the TRGB detection.  To provide the same distance, our observed F110W magnitude for the TRGB would be $\sim$22.1, near where there is a very modest increase in the stellar density (see Figure \ref{4panelTRGB}). On the other hand, our deep CMD reveals a clear discontinuity at m = 25.13 $\pm$ 0.03 in F110W, with a modest AGB population occupying the $\sim$1.5 magnitudes above (see EDD for a full-field CMD). The stars located near the previously reported TRGB magnitude of K=$19.90 \pm 0.09$, are instead either foreground stars, the brightest stars associated with the galaxy itself, or a combination of both.

It was pointed out by \cite{2003AJ....126.2362M} that the observed gas-phase metallicity ([O/H] = -0.9) in the singular HII region is significantly lower than the metallicity of the observed RGB ([Fe/H] = -0.5). \cite{2005MNRAS.359L..53B} resolve this issue by presenting Giant Metrewave Radio Telescope (GMRT) observations that resolve the galaxy into two separate components, with HIZSS-003A being much larger than the HII region hosting companion (HIZSS-003B). Our HST imaging reveals this smaller companion galaxy to be $1.4^{\prime}$ away.  The stellar population of the companion is too scant to derive an independent distance but the degree of resolution is comparable in the two systems. Given the small velocity difference ($\sim$35 \kms) and disturbed kinematical nature of the system presented with the GMRT data, the two almost certainly form an interacting system $\sim 3$~kpc apart.

\citet{2005MNRAS.359L..53B} determined HI and dynamical masses for the individual galaxies.  Adjusting the physical size of the system for our larger distance ($\sim$4$\times$), we find the HI and total masses of HIZSS-003A to be $2.2\times10^8~M_{\odot}$ and $2.6\times10^9~M_{\odot}$ respectively and, correspondingly for system B, $4.0\times10^7~M_{\odot}$ and $2\times10^8~M_{\odot}$.

\begin{figure*}[t!]
\centering
\gridline{\fig{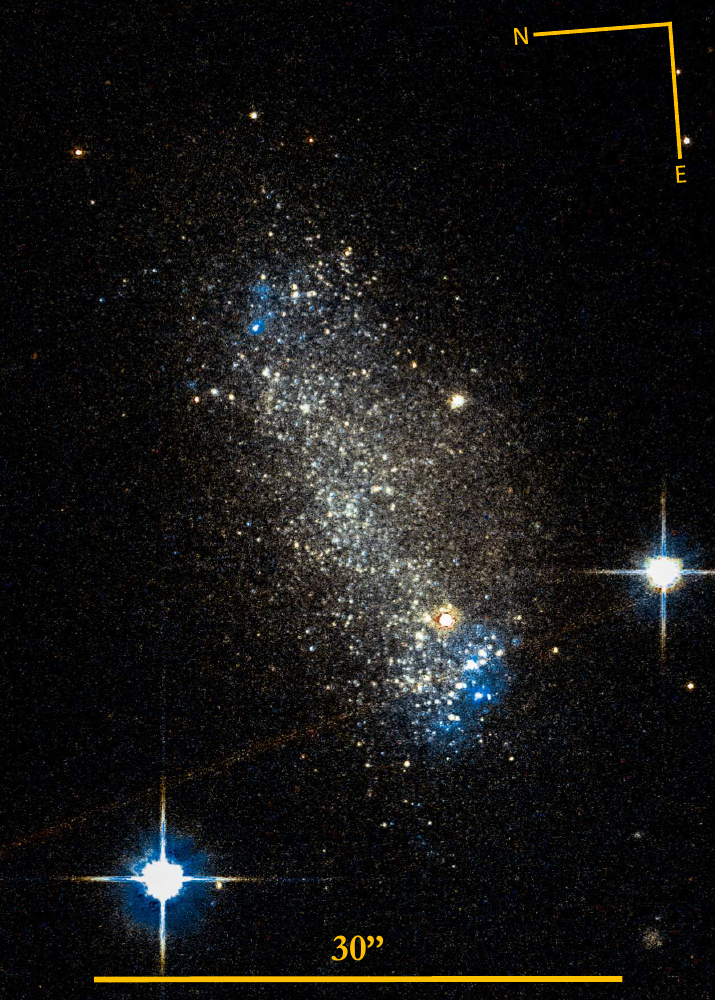}{0.5\textwidth}{(a)}
          \fig{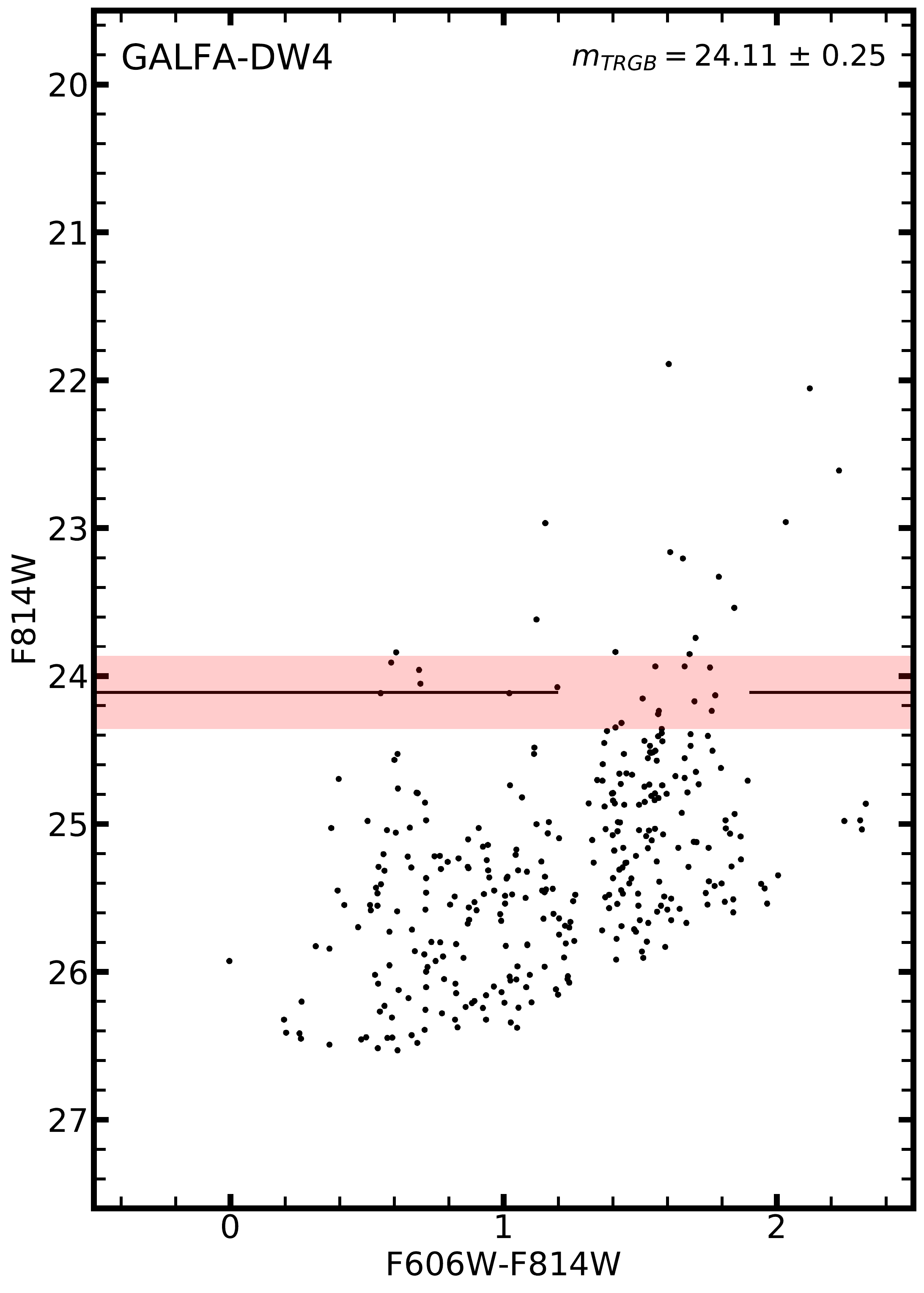}{0.5\textwidth}{(b)}}
\caption{\textbf{a:} Color image of GALFA-DW4 created from HST/ACS observations. F606W is mapped to blue, F814W is mapped to red, and a combination of the two is used for green. \textbf{b:} Color-magnitude diagram used for determining the distance to GALFA-DW4. The TRGB magnitude is relatively uncertain due to the sparseness of what is taken to be the RGB.}
\label{GALFA-DW4}
\end{figure*}

\subsubsection{Orion Dwarf}

The Orion Dwarf was first detected in HI by \cite{1979ApJ...227L.125G} using the Green Bank and Arecibo telescopes. Optical imaging from the BTA 6m telescope was used to determine the magnitude of the brightest stars, which led to a distance of 6.4 $\pm$ 2.0 Mpc \citep{1996A&A...315..348K}. \cite{2005AJ....130.1593V} later used infrared observations and a fundamental-plane defined by absolute magnitude, surface brightness, and HI line width to derive a distance of 5.4 $\pm$ 1.0 Mpc. \cite{2010AJ....139.2170C} present optical (Kitt Peak 0.9m) and HI (VLA) observations, which they use to characterize the rotation curve and star formation in the galaxy. They use H$\alpha$ imaging to derive a star formation rate of 0.04 $M_\odot/yr$. Their observed rotation curve flattens out at $\sim$ 80 \kms, and remains flat until their detection limit. From their measurements, they find that Orion is a relatively dark matter dominated system, with M/L $\approx$ 13.

\cite{2012MNRAS.426..751F} examine the rotation curve of Orion further, and attempt to consolidate the data with a standard formalism of Modified Newtonian Dynamics (MOND). Their Figures 7+8 show the result of this experiment. When assuming a distance of 5.4 Mpc \citep{2010AJ....139.2170C}, their adopted MOND formalism \citep{2005MNRAS.363..603F} does a poor job at modeling the observed rotation curve. Instead, their best fit is achieved by shifting the distance out to 10 Mpc. They highlight the need for a direct distance measurement via the near-infrared TRGB, which we now provide. Our distance of 6.79 $\pm$ 0.57 Mpc disfavors the MOND interpretation. \newline

\begin{table*}[t!]
\centering
\begin{tabular}{|c|c|c|c|c|c|c|c|}
\hline
\textbf{PGC} & \textbf{Galaxy} & \textbf{E(B-V)}  & \textbf{Distance} & \boldmath{\textbf{$v_{h}$}} & \boldmath{\textbf{$v_{LS}$}} & \textbf{$H_{0}d$} & \boldmath{\textbf{$v_{pec}$}}  \\ \hline
20171        & ESO 558-011     & 0.395 $\pm$ 0.04 & 7.16 $\pm$ 0.55  & 731 & 491         & 537                     & $-46$  $\pm$ 43                     \\ \hline
2807061      & HIZSS-003       & 0.845 $\pm$ 0.06 & 6.65 $\pm$ 0.50  & 288 & 115         & 499                     & $-384$ $\pm$ 40                     \\ \hline
2807065      & HIZSS-021       & 0.767 $\pm$ 0.04 & 5.78 $\pm$ 0.39  & 494 & 220         & 434                     & $-214$ $\pm$ 32                     \\ \hline
17716        & Orion Dwarf     & 0.595 $\pm$ 0.05 & 6.79 $\pm$ 0.57  & 387 & 303         & 509                     & $-206$ $\pm$ 45                     \\ \hline
5072715      & GALFA-DW4       & 0.530 $\pm$ 0.05  & 2.98 $\pm$ 0.37 & 630 & 572         & 224                     & +348   $\pm$ 28                     \\ \hline
\end{tabular}
\caption{A summary of our distances results for the five targets discussed in $\S$3, including PGC Catalog $\#$, derived reddening, distance (Mpc), heliocentric velocity ($v_{h}$), Local Sheet velocity ($v_{LS}$), expected Hubble flow velocity ($H_{0}d$) assuming $H_{0}$ = 75 \kmsMpc, and peculiar velocity ($v_{pec} = v_{LS}-H_{0}d$). All velocities are given in \kms.}
\label{distances}
\end{table*}

\subsection{The Case of GALFA-DW4}

There is another galaxy located near the south supergalactic pole whose distance we were able to evaluate during the process of our archival work ($\S$2.1). This galaxy, GALFA-DW4, was found during a search for optical counterparts to ultra-compact high velocity clouds by \cite{2015ApJ...806...95S}. They found a galactic counterpart to the Galactic Arecibo L-Band Feed Array HI \citep{2006ApJ...653.1210S} source GALFA-HI 086.4+10.8+611 at J2000 05:45:44.8 +10:46:16 (Galactic $l=195.6610~b=-09.3213$, supergalactic $sgl=352.4686~sgb=-58.4543$). This galaxy is $5.7^{\circ}$ from the Orion dwarf in projection. The velocity of the HI signal ($v_{hel}$=629.6 \kms) is consistent with the H$\alpha$ velocity of the optical counterpart ($v_{hel}$=622$\pm$35 \kms). It turns out that our derived distance to GALFA-DW4 indicates that it has a substantial positive peculiar velocity (see $\S$4).  Here we provide details about the analysis of this interesting case.

GALFA-DW4 was observed with HST/ACS during Cycle 24 (HST Proposal 14676, PI D. Sand) with one orbit split evenly between F606W $\&$ F814W.  We used this data to generate a CMD and derive a TRGB distance in the course of our standard archival analysis.  For the TRGB determination, in addition to restricting the field to the galaxy to diminish foreground contamination, we opt to exclude two prominent regions of star-formation located at each ends of the galaxy (see left-panel of Figure \ref{GALFA-DW4}). These regions are the sites of supergiants which have similar magnitudes and colors to stars at the TRGB, and excluding them decreases confusion from young populations. 

The CMD (right-panel of Figure \ref{GALFA-DW4}) is quite sparse, making it difficult with our maximum likelihood methodology to determine the magnitude of the TRGB \citep{2009ApJ...690..389M}. More crudely, we use a Sobel filter to determine $m_{TRGB} = 24.11 \pm 0.25$, with the generous uncertainty given by two alternate responses from the Sobel filter. The case for the brighter magnitude depends on whether those few stars at m $\sim23.85$ are AGB stars or part of the RGB. On the lower side, a reasonable case can be made that the RGB reaches at least ($\sim24.35$). Adopting E(B-V) = 0.53 from \cite{2011ApJ...737..103S}, we find a distance to GALFA-DW4 of 2.98 $\pm$ 0.37 Mpc. The error in the distance ($\sim$13$\%$) is larger than our typical uncertainties for ACS observations ($\sim$5$\%$), but is compelled by the relative sparseness of what we have taken to be the RGB.  A summary of distance and velocity information is provided in Table \ref{distances}. \newline \newline

\begin{figure*}[t!]
\centering
\gridline{\fig{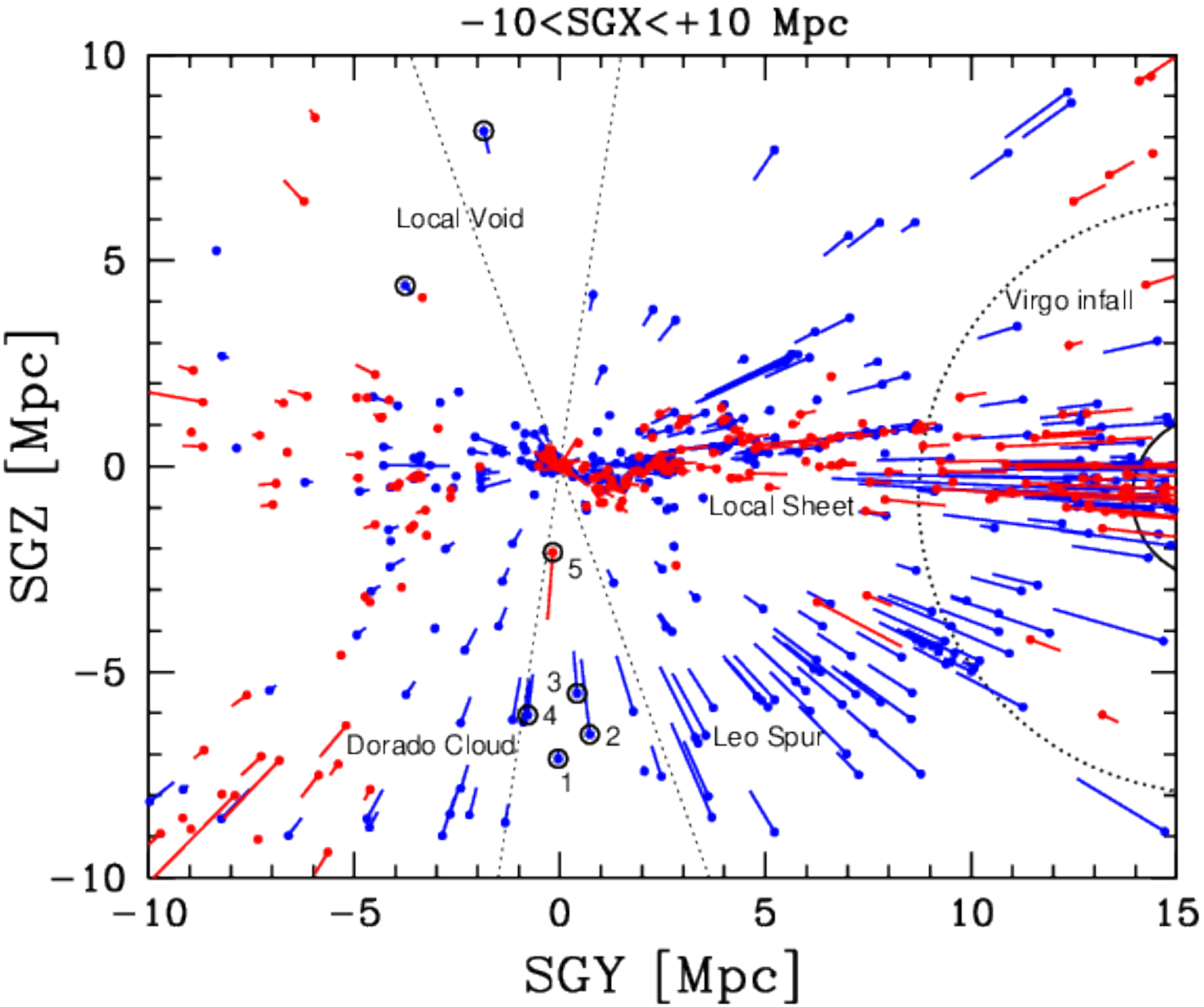}{0.5\textwidth}{(a)}
          \fig{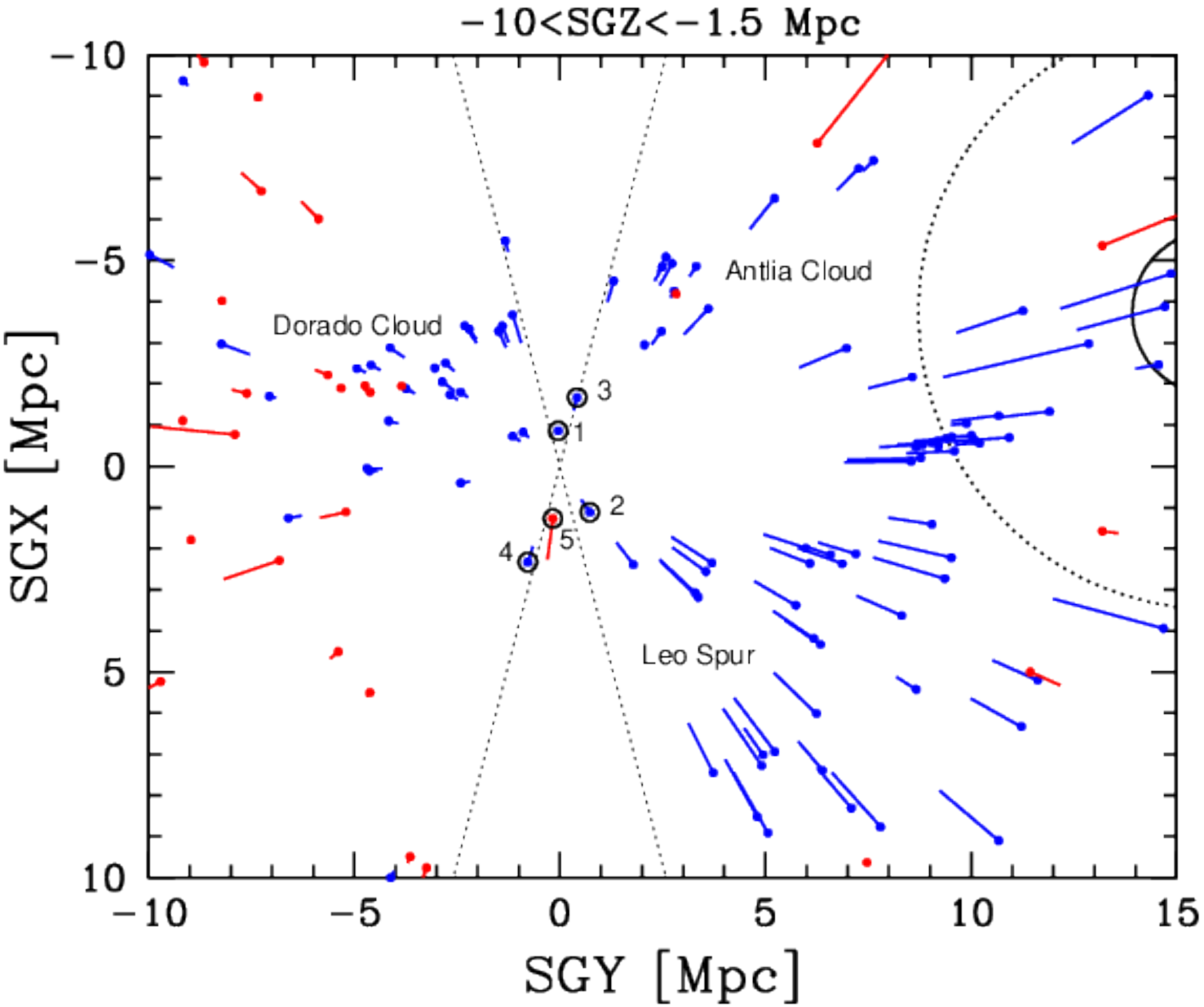}{0.5\textwidth}{(b)}}
\caption{Projected negative peculiar velocities (blue) and positive peculiar velocities (red).  Solid and dotted circles delineate the Virgo Cluster 2nd and 1st turnaround radii.  The dotted wedge delineates $\vert b \vert < 15^{\circ}$. Names are given to structural features.  The 5 galaxies of particular interest are (1) ESO 558-011, (2) HIZSS-003, (3) HIZSS-021, (4) Orion Dwarf, (5) GALFA-DW4.  \textbf{a:} Projection edge-on to the Local Sheet on the supergalactic equatorial plane. \textbf{b:} Structure at SGZ$<-1.5$~Mpc below the equatorial plane viewed from the supergalactic pole.}
\label{xyz}
\end{figure*}


\section{Peculiar Velocities}

This investigation was motivated by the patterns of motions seen in Figure~\ref{xyz}.  Accurate distance measurements allow discrimination of deviant motions from cosmic expansion.  Observed velocities, $V_{obs}$, can be separated between the expansion component, $H_0d$, and the line-of-sight peculiar velocity, $V_{pec}$,
\begin{equation}
    V_{obs} = H_od + V_{pec}.
\end{equation}
The appropriate reference frame for the analysis of velocities in the domain $1-15$ Mpc is with respect to the Local Group, and here we use the ``Local Sheet" variant \citep{2008ApJ...676..184T}.  The choice of Hubble constant must be consistent with the scaling of the distance estimators, which for this sample is $H_0=75$~\kmsMpc\ \citep{2016AJ....152...50T}.  

The vectors seen in Figure~\ref{xyz} represent projections of $V_{pec}$, blue for negative values and red for positive values, emanating from the dots at the measured positions of the target galaxies.  Only galaxies with errors on distance less than 15\% are included. 
Most galaxies in the volume projected into these figures reside either within $\pm2$~Mpc of the equatorial plane of the supergalactic coordinate system (the Local Sheet) or, below the equatorial plane, in the sparse, interconnected filaments called Leo Spur, Dorado Cloud, and Antlia Cloud in the Nearby Galaxies Atlas \citep{1987nga..book.....T}. 

In the projection edge-on to these features, the left panel of Figure~\ref{xyz}, it is found that motions within the Local Sheet in the interval $-5<SGY<+8$~Mpc are modest and scrambled about Hubble flow expectations.  Beyond $SGY \simeq +8$~Mpc peculiar velocities become systematically positive, manifesting infall toward the Virgo Cluster \citep{2014ApJ...782....4K,2018ApJ...858...62K}.  Positive residuals at negative $SGY$ mark the approach to the Fornax Cluster and the structure \cite{1956VA......2.1584D} called the Southern Supergalaxy, lying to the left of the display of panel $a$. 

There are few galaxies above the supergalactic equatorial plane (positive $SGZ$), the domain of the Local Void. Some of these lie along the M101 Wall, a spur off the Local Sheet extending into the Local Void \citep{2017A&A...602A.119M,2018AJ....156..105A}. This feature is visible at positive SGY, and extends up to SGZ $\sim$+6. Those galaxies lying deeper within the void that have been studied in detail have peculiar velocities toward us. We interpret these motions as due to the expansion of the Local Void \citep{2017ApJ...835...78R}. Galaxies on the far side of the Local Void should also be experiencing the effects of the void's expansion, though accurate peculiar velocities are difficult to obtain at such large distances \citep{2011A&A...531A..87I}.

Our particular interest with the present study is in the structures {\it below} the supergalactic equatorial plane that overwhelmingly are also found to have velocities toward us (at $SGY < -5$~Mpc there is a transition to $V_{pec}$ positive as the Fornax Cluster is approached).  From our standpoint in co-moving coordinates, essentially all nearby galaxies above and below the Local Sheet ($\vert SGZ \vert > 2$~Mpc) are heading toward us. 

While the pattern of {\it relative} motions was becoming apparent, the {\it absolute} motions with regard to very large scales was not secure.  Specifically, we can expect that our Local Sheet is moving toward negative $SGZ$ due to the expansion of the Local Void, but with what amplitude?  To what degree are the negative velocities of galaxies at negative $SGZ$ just a reflex of our downward motion versus a possible consequence of their upward motion?  

To a reasonable approximation, the answer to this question is known. \citet{2017ApJ...850..207S} reconstructed the orbits of 1382 halos that capture the distribution of mass within 38~Mpc with numerical action methods.  The reconstruction was done in the reference frame of the ensemble of mass within this volume.  Overwhelmingly, the reconstructed orbits of galaxies above and within the supergalactic equatorial plane have orbits toward negative $SGZ$ in co-moving coordinates.  By contrast, galaxies below the structures on the equatorial plane have neutral velocities in the $SGZ$ direction.  In the orthogonal directions, both within the Local Sheet and below the equatorial plane, galaxies are systematically flowing toward positive $SGY$ (toward the Virgo Cluster), and toward negative $SGX$ (toward the CMB dipole apex).

\begin{figure*}[t!]
\centering
\gridline{\fig{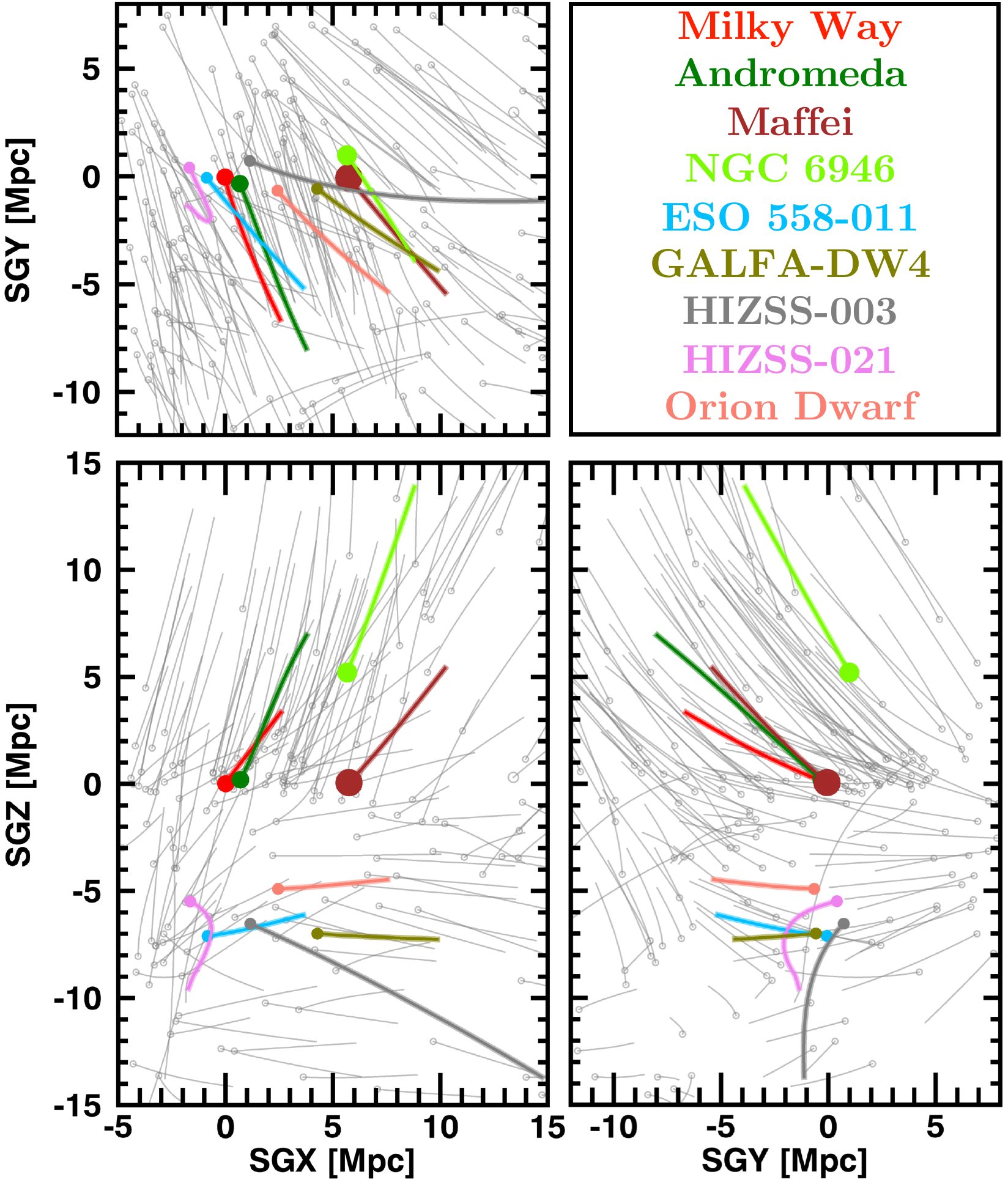}{1\textwidth}{}}
\caption{Numerical action methods orbits for the galaxies being discussed in diverse colors (additional galaxies in grey) shown from three perspectives. All galaxies participate in the streaming pattern seen in the projection normal to the SGZ axis. Galaxies on and above the supergalactic equator (SGZ$>0$) all have downward motions, a characteristic not shared by galaxies at negative SGZ. This downward motion is a direct result of the expansion of the Local Void, resulting in the piling up of galaxies onto the Local Sheet.}
\label{orbit}
\end{figure*}

The intent of the new observations was to provide confirmation of the most local of these motions, our motion normal to the equatorial plane and away from the Local Void.  Line-of-sight peculiar velocities in the $SGZ$ direction are maximized toward the supergalactic pole. It is seen in Figure~\ref{xyz} that this direction is toward the zone of obscuration.  

The four targets of our program are found to lie roughly as anticipated in structures $\sim6.5$~Mpc below the equatorial plane.  Those labeled 1 and 3 in the figure are at the interface between the Antlia Cloud and Dorado Cloud features (ESO 558-011 with $V_{pec}=-46\pm43$~\kms\ and HIZSS-021 with $V_{pec}=-214\pm32$~\kms).  Those labeled 2 and 4 (HIZSS-03 with $V_{pec}=-374\pm40$~\kms\ and Orion dwarf with $V_{pec}=-223\pm45$~\kms)) are associated with the Leo Spur near the junction with Dorado Cloud. Here the errors reflect the uncertainties in distance convolved with uncertainty in the proper value of $H_0$ for the problem.  The latter uncertainty is small: $\Delta H_0 d \sim 12 - 14$~\kms\ with $\Delta H_0 = 2$~\kmsMpc.  Three of these galaxies clearly have negative $V_{pec}$ and the fourth is negative with $1\sigma$ significance.

The newly derived distances were embedded in the \citet{2017ApJ...850..207S} model and the resultant orbits (from z=4 to the present) are shown in Figure~\ref{orbit}. In the the SGX$-$SGY projection, there is great coherence. All these galaxies are flowing toward positive SGY (Virgo Cluster) and negative SGX (Great Attractor, apex CMB dipole). In the SGX$-$SGZ and SGY-SGZ projections, the galaxies on and above the supergalactic equator (positive SGZ) all have motions that are descending in SGZ. Those at negative SGZ, in the Leo Spur$-$Dorado Cloud filament, have scrambled SGZ motions.

As incidentals, NGC~6946 and the Maffei galaxy are plotted because new distances have become available for these two dynamically important players \citep{2018AJ....156..105A,2019ApJ...872L...4A}.  Their orbits follow the now familiar patterns.  If it is considered curious that all the galaxies being plotted lie at SGY$\sim 0$, it is because the galactic plane has this value.  All the galaxies being discussed lie at low galactic latitude.

\begin{figure*}
\centering
\gridline{\fig{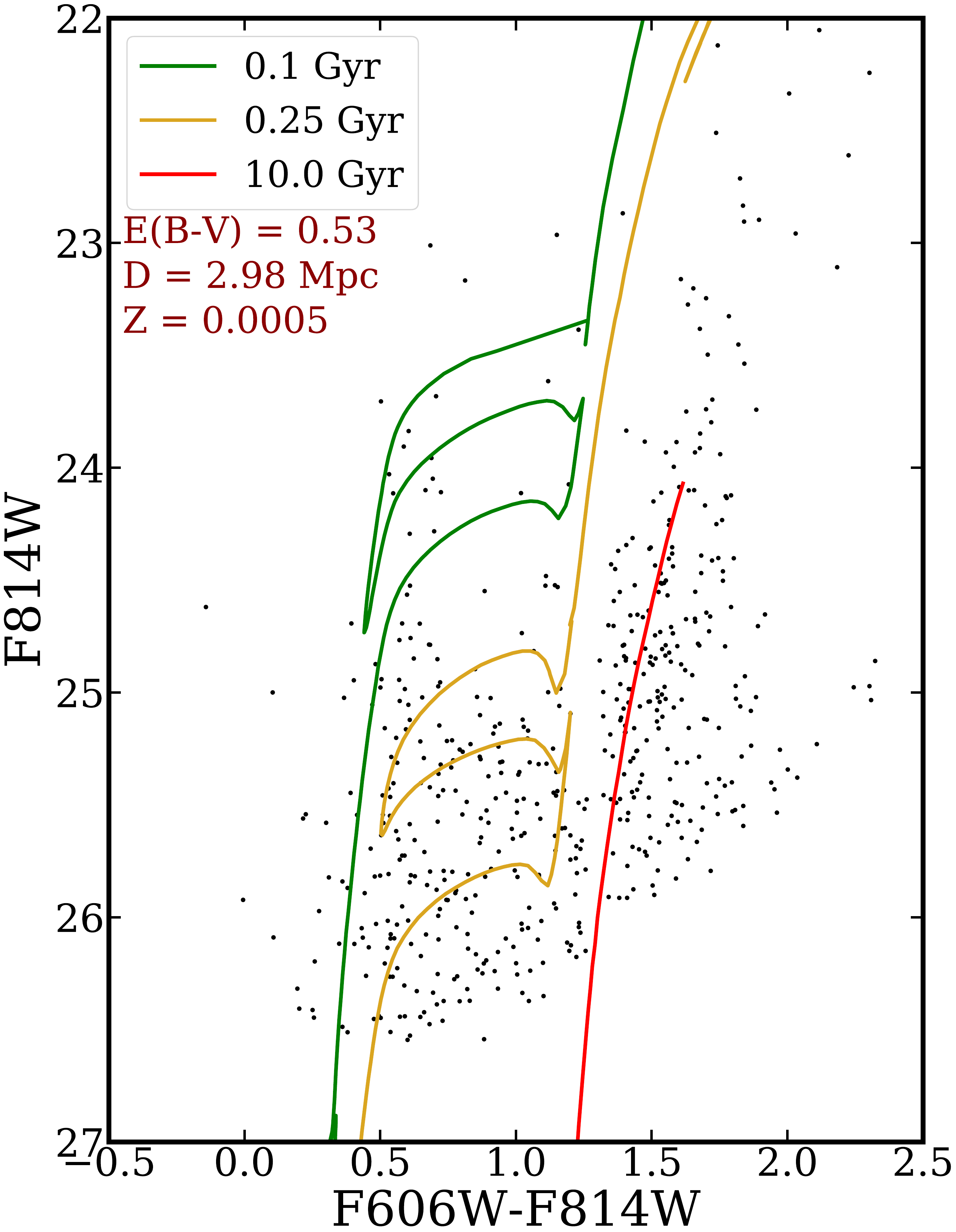}{0.5\textwidth}{(a)}
          \fig{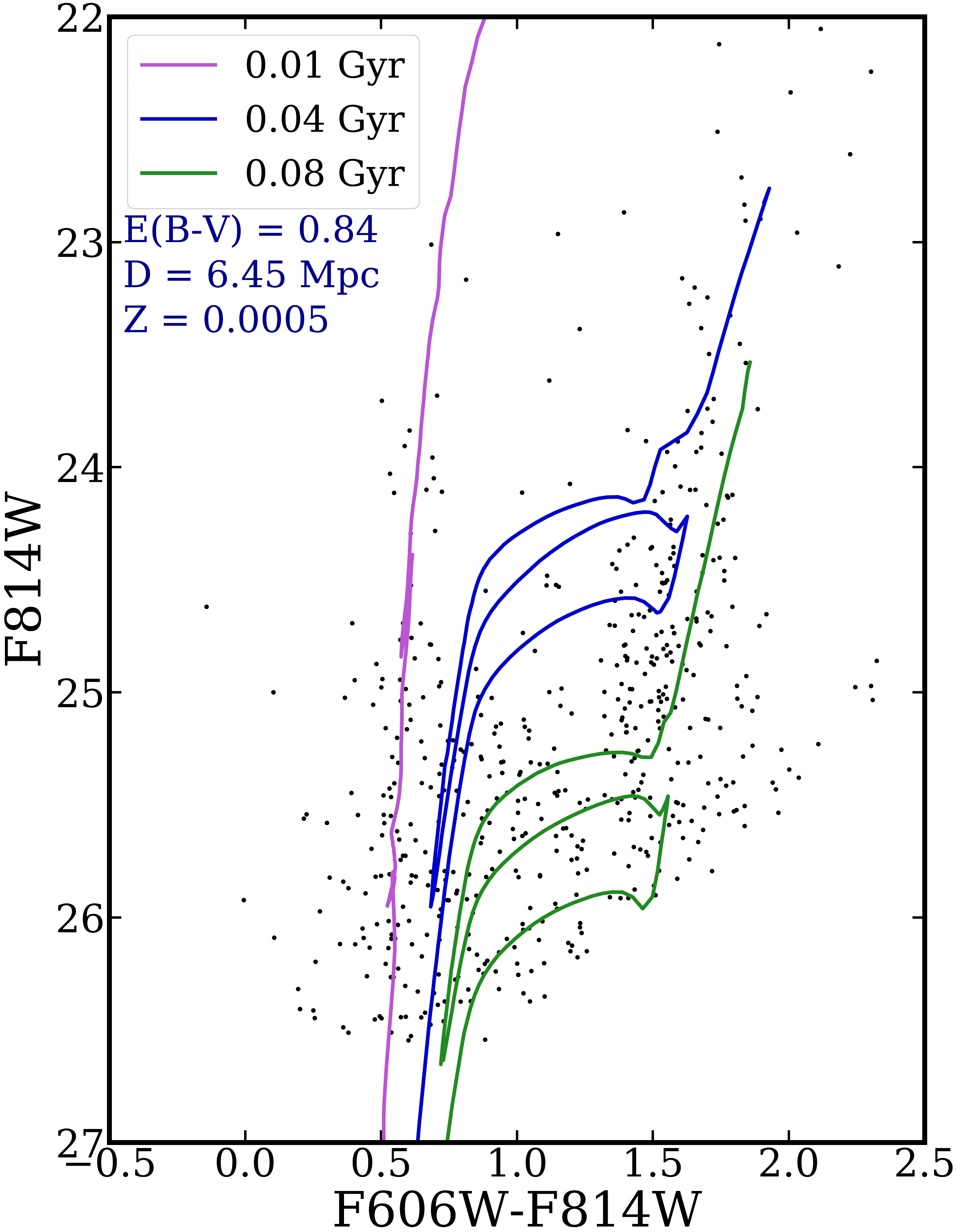}{0.5\textwidth}{(b)}}
\caption{Two distance options for GALFA-DW4. \textbf{a:} At 3~Mpc the RGB reaches almost F814W=24.  Bluer stars and those brighter than 24 mag would have ages of a few hundred Myr. \textbf{b:} At greater than twice the distance, the TRGB would lie fainter than 26 mag.  The resolved stars would all have ages less than 100 Myr.}
\label{isochrone}
\end{figure*}

\subsection{GALFA-DW4: The Near Scenario}
 The fifth object that came to our attention, GALFA-DW4, which lies in close proximity in projection to the Orion dwarf, is a puzzle because at its measured distance of 3.0~Mpc the implied peculiar motion is $V_{pec}=+348\pm28$~\kms. The uncertainty due to choice of $H_0$ is an insignificant 6~\kms.  Though what is taken to be the red giant branch is sparsely populated, it is sufficiently dense by F814W$\sim 24.4$ that the system would not be more distant than 3.4~Mpc, whence the peculiar velocity would not be less than $\sim +320$~\kms.

One potential explanation for this velocity anomaly is that GALFA-DW4 is not truly isolated; that there is a large galaxy lurking in the vicinity influencing its motion. While unlikely, the enhanced extinction in this region due to the proximity of the galactic plane raises the small possibility that such a galaxy may have been missed in previous catalogs. A gas-rich target would result in a large HI signal, and thus a significant detection in a survey such as GALFA. The other option is that this hypothetical galaxy is instead a gas-poor elliptical, and has been incorrectly catalogued in existing all-sky surveys.

This possibility has been examined by giving attention to all sources in a 460 deg$^{2}$ region around GALFA-DW4 in the 2MASS Extended Source Catalog \citep{2006AJ....131.1163S}.  $J$+$H$+$K_{s}$ color cutouts allow us to exclude clearly distinguishable galactic nebulae and young stellar objects. All identified galaxies are much less intrinsically luminous than expected for a large elliptical at the required distance of $\sim$3 Mpc. It is conceivable that the hypothetical target is located directly behind a very bright star in the region (such as Betelgeuse, $l=199.7872, b=-8.9586$), which would make it difficult to clearly detect. 

\subsection{GALFA-DW4: The Far Scenario}

In the left panel of Figure~\ref{isochrone} PARSEC stellar population isochrones \citep{2012MNRAS.427..127B,2017ApJ...835...77M,2019MNRAS.tmp..738P} are imposed on the CMD of GALFA-DW4 assuming a distance of 2.98 Mpc.  The RGB is represented by a 10 Gyr isochrone with very low metallicity that terminates at the TRGB.  Stars blueward of the RGB and the modest number of stars above the RGB are described by stellar populations of a few hundred Myr.

The right panel presents an alternative.  Here, assuming a distance of 6.5~Mpc, all the resolved stars lie above the RGB.  Their ages are 80 Myr and less, with a component of ongoing star formation consistent with the observed emission regions.  The stars taken to be on the red giant branch at a near distance would actually be on the asymptotic giant branch at a far distance.  The TRGB would be lurking faintward of F814W$\sim 26$.  

The assumed reddening is different in the two cases because at the farther distance the bluest stars are on the zero age main sequence whereas at the nearer distance the bluest resolved stars have evolved off the zero age line. In the farther case, we use the same Sobel filtering method we performed with Orion to find E(B-V) = 0.84 for the GALFA-DW4 dwarf.

The true situation cannot be resolved with the information currently available.  A deeper CMD would resolve the matter.  Either there would be a break in the stellar luminosity function below F814W$\sim26$, the manifestation of the onset of the RGB, or no break, the expectation if more sensitive observations simply reveal less luminous stars on the RGB than those now seen.

An orbit for GALFA-DW4 has been plotted in Figure~\ref{orbit}. In lieu of having a distance, the numerical action model found an orbit with an end point consistent with the observed velocity. A distance of $\sim 7$~Mpc is obtained. Without a mystery attractor, no plausible orbit for GALFA-DW4 within the model could end up at 3 Mpc.

\section{Summary}

Four modest galaxies lying in the zone of obscuration have been observed with HST WFC3 at infrared bands to successfully resolve stars within the red giant branch.  Distances are determined from the luminosities of stars at the tips of the red giant branch. In all cases, the distances are greater than they would be given their observed velocities and the assumption of participation in the Hubble expansion.  They all have negative velocities in co-moving coordinates.

This result was largely anticipated.  The newly observed galaxies lie in a bridge between filaments called the Leo Spur north of the galactic plane and the Doradus Cloud south of the plane.  Galaxies across this continuity of structure systematically manifest negative peculiar velocities.  The motions have been interpreted in the past as a reflex of the velocities of our Galaxy and immediate neighbors in the Local Sheet away from the Local Void.  The Local Sheet is part of a wall of an expanding void, with the expansion pushing us toward the Leo Spur $-$ Doradus Cloud filament.

An additional galaxy close by on the sky observed in a separate HST program, GALFA-DW4, came to our attention.  It would appear from the CMD that the RGB is being resolved which would place the galaxy nearby, at 3.0~Mpc, and with a substantial {\it positive} peculiar velocity of 348~\kms.  This situation would be very unusual and interesting, if true.  Alternatively, the feature that we are taking to be the red giant branch might be the asymptotic giant branch.  The red giant branch might be unresolved.  With this explanation, GALFA-DW4 lies at greater than 6 Mpc and has a peculiar velocity not in substantial disagreement with the other systems in this program.

\bigskip

This research is supported by an award from the Space Telescope Science Institute in support of program GO-15150. I.K. acknowledges support by RFBR grant 18-02-00005. We thank the anonymous referee for their useful comments that helped improve our manuscript. G.A. would like to thank C. Auge, Z. Claytor, A. Payne, $\&$ M. Tucker for useful discussions throughout the length of the paper. 

\facilities{HST (ACS/WFC, WFC3/IR)}

\software{APLpy \citep{2012ascl.soft08017R}, DOLPHOT \citep{2000PASP..112.1397D,2016ascl.soft08013D}, DrizzlePac 2.0 \citep{2015ASPC..495..281A}, Montage \citep{2010ascl.soft10036J, 2017ASPC..512...81B},  TRILEGAL \citep{2005A&A...436..895G}}


\bibliography{paper}

\begin{thebibliography}{}
\expandafter\ifx\csname natexlab\endcsname\relax\def\natexlab#1{#1}\fi
\providecommand{\url}[1]{\href{#1}{#1}}

\bibitem[{{Anand} {et~al.}(2018){Anand}, {Rizzi}, \&
  {Tully}}]{2018AJ....156..105A}
{Anand}, G.~S., {Rizzi}, L., \& {Tully}, R.~B. 2018, \aj, 156, 105

\bibitem[{{Anand} {et~al.}(2019){Anand}, {Tully}, {Rizzi}, \&
  {Karachentsev}}]{2019ApJ...872L...4A}
{Anand}, G.~S., {Tully}, R.~B., {Rizzi}, L., \& {Karachentsev}, I.~D. 2019,
  \apj, 872, L4

\bibitem[{{Avila} {et~al.}(2015){Avila}, {Hack}, {Cara}, {Borncamp}, {Mack},
  {Smith}, \& {Ubeda}}]{2015ASPC..495..281A}
{Avila}, R.~J., {Hack}, W., {Cara}, M., {et~al.} 2015, in Astronomical Society
  of the Pacific Conference Series, Vol. 495, Astronomical Data Analysis
  Software an Systems XXIV (ADASS XXIV), ed. A.~R. {Taylor} \& E.~{Rosolowsky},
  281

\bibitem[{{Beaton} {et~al.}(2018){Beaton}, {Bono}, {Braga}, {Dall'Ora},
  {Fiorentino}, {Jang}, {Mart{\'\i}nez-V{\'a}zquez}, {Matsunaga}, {Monelli},
  {Neeley}, \& {Salaris}}]{2018SSRv..214..113B}
{Beaton}, R.~L., {Bono}, G., {Braga}, V.~F., {et~al.} 2018, \ssr, 214, 113

\bibitem[{{Begum} {et~al.}(2005){Begum}, {Chengalur}, {Karachentsev}, \&
  {Sharina}}]{2005MNRAS.359L..53B}
{Begum}, A., {Chengalur}, J.~N., {Karachentsev}, I.~D., \& {Sharina}, M.~E.
  2005, \mnras, 359, L53

\bibitem[{{Berriman} {et~al.}(2017){Berriman}, {Good}, {Rusholme}, \&
  {Robitaille}}]{2017ASPC..512...81B}
{Berriman}, G.~B., {Good}, J.~C., {Rusholme}, B., \& {Robitaille}, T. 2017, in
  Astronomical Society of the Pacific Conference Series, Vol. 512, Astronomical
  Data Analysis Software and Systems XXV, ed. N.~P.~F. {Lorente},
  K.~{Shortridge}, \& R.~{Wayth}, 81

\bibitem[{{Bressan} {et~al.}(2012){Bressan}, {Marigo}, {Girardi}, {Salasnich},
  {Dal Cero}, {Rubele}, \& {Nanni}}]{2012MNRAS.427..127B}
{Bressan}, A., {Marigo}, P., {Girardi}, L., {et~al.} 2012, \mnras, 427, 127

\bibitem[{{Cannon} {et~al.}(2010){Cannon}, {Haynes}, {Most}, {Salzer},
  {Haugland}, {Scudder}, {Sugden}, \& {Weindling}}]{2010AJ....139.2170C}
{Cannon}, J.~M., {Haynes}, K., {Most}, H., {et~al.} 2010, \aj, 139, 2170

\bibitem[{{Da Costa} \& {Armandroff}(1990)}]{1990AJ....100..162D}
{Da Costa}, G.~S., \& {Armandroff}, T.~E. 1990, \aj, 100, 162

\bibitem[{{Dalcanton} {et~al.}(2012){Dalcanton}, {Williams}, {Melbourne},
  {Girardi}, {Dolphin}, {Rosenfield}, {Boyer}, {de Jong}, {Gilbert}, {Marigo},
  {Olsen}, {Seth}, \& {Skillman}}]{2012ApJS..198....6D}
{Dalcanton}, J.~J., {Williams}, B.~F., {Melbourne}, J.~L., {et~al.} 2012,
  \apjs, 198, 6

\bibitem[{{de Vaucouleurs}(1956)}]{1956VA......2.1584D}
{de Vaucouleurs}, G. 1956, Vistas in Astronomy, 2, 1584

\bibitem[{{Dolphin}(2016)}]{2016ascl.soft08013D}
{Dolphin}, A. 2016, {DOLPHOT: Stellar photometry}, , , ascl:1608.013

\bibitem[{{Dolphin}(2000{\natexlab{a}})}]{2000PASP..112.1383D}
{Dolphin}, A.~E. 2000{\natexlab{a}}, \pasp, 112, 1383

\bibitem[{{Dolphin}(2000{\natexlab{b}})}]{2000PASP..112.1397D}
---. 2000{\natexlab{b}}, \pasp, 112, 1397

\bibitem[{{Famaey} \& {Binney}(2005)}]{2005MNRAS.363..603F}
{Famaey}, B., \& {Binney}, J. 2005, \mnras, 363, 603

\bibitem[{{Frusciante} {et~al.}(2012){Frusciante}, {Salucci}, {Vernieri},
  {Cannon}, \& {Elson}}]{2012MNRAS.426..751F}
{Frusciante}, N., {Salucci}, P., {Vernieri}, D., {Cannon}, J.~M., \& {Elson},
  E.~C. 2012, \mnras, 426, 751

\bibitem[{{Giovanelli}(1979)}]{1979ApJ...227L.125G}
{Giovanelli}, R. 1979, \apjl, 227, L125

\bibitem[{{Girardi} {et~al.}(2005){Girardi}, {Groenewegen}, {Hatziminaoglou},
  \& {da Costa}}]{2005A&A...436..895G}
{Girardi}, L., {Groenewegen}, M.~A.~T., {Hatziminaoglou}, E., \& {da Costa}, L.
  2005, \aap, 436, 895

\bibitem[{{Henning} {et~al.}(2000){Henning}, {Staveley-Smith}, {Ekers},
  {Green}, {Haynes}, {Juraszek}, {Kesteven}, {Koribalski}, {Kraan-Korteweg},
  {Price}, {Sadler}, \& {Schr{\"o}der}}]{2000AJ....119.2686H}
{Henning}, P.~A., {Staveley-Smith}, L., {Ekers}, R.~D., {et~al.} 2000, \aj,
  119, 2686

\bibitem[{{Iwata} \& {Chamaraux}(2011)}]{2011A&A...531A..87I}
{Iwata}, I., \& {Chamaraux}, P. 2011, \aap, 531, A87

\bibitem[{{Jacob} {et~al.}(2010){Jacob}, {Katz}, {Berriman}, {Good}, {Laity},
  {Deelman}, {Kesselman}, {Singh}, {Su}, {Prince}, \&
  {Williams}}]{2010ascl.soft10036J}
{Jacob}, J.~C., {Katz}, D.~S., {Berriman}, G.~B., {et~al.} 2010, {Montage: An
  Astronomical Image Mosaicking Toolkit}, , , ascl:1010.036

\bibitem[{{Jacobs} {et~al.}(2009){Jacobs}, {Rizzi}, {Tully}, {Shaya},
  {Makarov}, \& {Makarova}}]{2009AJ....138..332J}
{Jacobs}, B.~A., {Rizzi}, L., {Tully}, R.~B., {et~al.} 2009, \aj, 138, 332

\bibitem[{{Karachentsev} \& {Musella}(1996)}]{1996A&A...315..348K}
{Karachentsev}, I., \& {Musella}, I. 1996, \aap, 315, 348

\bibitem[{{Karachentsev} {et~al.}(2018){Karachentsev}, {Makarova}, {Tully},
  {Rizzi}, \& {Shaya}}]{2018ApJ...858...62K}
{Karachentsev}, I.~D., {Makarova}, L.~N., {Tully}, R.~B., {Rizzi}, L., \&
  {Shaya}, E.~J. 2018, \apj, 858, 62

\bibitem[{{Karachentsev} {et~al.}(2015){Karachentsev}, {Tully}, {Makarova},
  {Makarov}, \& {Rizzi}}]{2015ApJ...805..144K}
{Karachentsev}, I.~D., {Tully}, R.~B., {Makarova}, L.~N., {Makarov}, D.~I., \&
  {Rizzi}, L. 2015, \apj, 805, 144

\bibitem[{{Karachentsev} {et~al.}(2014){Karachentsev}, {Tully}, {Wu}, {Shaya},
  \& {Dolphin}}]{2014ApJ...782....4K}
{Karachentsev}, I.~D., {Tully}, R.~B., {Wu}, P.-F., {Shaya}, E.~J., \&
  {Dolphin}, A.~E. 2014, \apj, 782, 4

\bibitem[{{Karachentsev} {et~al.}(2002){Karachentsev}, {Sharina}, {Makarov},
  {Dolphin}, {Grebel}, {Geisler}, {Guhathakurta}, {Hodge}, {Karachentseva},
  {Sarajedini}, \& {Seitzer}}]{2002A&A...389..812K}
{Karachentsev}, I.~D., {Sharina}, M.~E., {Makarov}, D.~I., {et~al.} 2002, \aap,
  389, 812

\bibitem[{{Karachentsev} {et~al.}(2003){Karachentsev}, {Makarov}, {Sharina},
  {Dolphin}, {Grebel}, {Geisler}, {Guhathakurta}, {Hodge}, {Karachentseva},
  {Sarajedini}, \& {Seitzer}}]{2003A&A...398..479K}
{Karachentsev}, I.~D., {Makarov}, D.~I., {Sharina}, M.~E., {et~al.} 2003, \aap,
  398, 479

\bibitem[{{Lee} {et~al.}(1993){Lee}, {Freedman}, \&
  {Madore}}]{1993ApJ...417..553L}
{Lee}, M.~G., {Freedman}, W.~L., \& {Madore}, B.~F. 1993, \apj, 417, 553

\bibitem[{{Madore} {et~al.}(2009){Madore}, {Mager}, \&
  {Freedman}}]{2009ApJ...690..389M}
{Madore}, B.~F., {Mager}, V., \& {Freedman}, W.~L. 2009, \apj, 690, 389

\bibitem[{{Makarov} {et~al.}(2006){Makarov}, {Makarova}, {Rizzi}, {Tully},
  {Dolphin}, {Sakai}, \& {Shaya}}]{2006AJ....132.2729M}
{Makarov}, D., {Makarova}, L., {Rizzi}, L., {et~al.} 2006, \aj, 132, 2729

\bibitem[{{Marigo} {et~al.}(2017){Marigo}, {Girardi}, {Bressan}, {Rosenfield},
  {Aringer}, {Chen}, {Dussin}, {Nanni}, {Pastorelli}, {Rodrigues}, {Trabucchi},
  {Bladh}, {Dalcanton}, {Groenewegen}, {Montalb{\'a}n}, \&
  {Wood}}]{2017ApJ...835...77M}
{Marigo}, P., {Girardi}, L., {Bressan}, A., {et~al.} 2017, \apj, 835, 77

\bibitem[{{Massey} {et~al.}(2003){Massey}, {Henning}, \&
  {Kraan-Korteweg}}]{2003AJ....126.2362M}
{Massey}, P., {Henning}, P.~A., \& {Kraan-Korteweg}, R.~C. 2003, \aj, 126, 2362

\bibitem[{{McQuinn} {et~al.}(2017){McQuinn}, {Skillman}, {Dolphin}, {Berg}, \&
  {Kennicutt}}]{2017AJ....154...51M}
{McQuinn}, K. B.~W., {Skillman}, E.~D., {Dolphin}, A.~E., {Berg}, D., \&
  {Kennicutt}, R. 2017, \aj, 154, 51

\bibitem[{{M{\"u}ller} {et~al.}(2017){M{\"u}ller}, {Scalera}, {Binggeli}, \&
  {Jerjen}}]{2017A&A...602A.119M}
{M{\"u}ller}, O., {Scalera}, R., {Binggeli}, B., \& {Jerjen}, H. 2017, \aap,
  602, A119

\bibitem[{{Pastorelli} {et~al.}(2019){Pastorelli}, {Marigo}, {Girardi}, {Chen},
  {Rubele}, {Trabucchi}, {Aringer}, {Bladh}, {Bressan}, {Montalb{\'a}n},
  {Boyer}, {Dalcanton}, {Eriksson}, {Groenewegen}, {H{\"o}fner}, {Lebzelter},
  {Nanni}, {Rosenfield}, {Wood}, \& {Cioni}}]{2019MNRAS.tmp..738P}
{Pastorelli}, G., {Marigo}, P., {Girardi}, L., {et~al.} 2019, \mnras, 738

\bibitem[{{Rizzi} {et~al.}(2007){Rizzi}, {Tully}, {Makarov}, {Makarova},
  {Dolphin}, {Sakai}, \& {Shaya}}]{2007ApJ...661..815R}
{Rizzi}, L., {Tully}, R.~B., {Makarov}, D., {et~al.} 2007, \apj, 661, 815

\bibitem[{{Rizzi} {et~al.}(2017){Rizzi}, {Tully}, {Shaya}, {Kourkchi}, \&
  {Karachentsev}}]{2017ApJ...835...78R}
{Rizzi}, L., {Tully}, R.~B., {Shaya}, E.~J., {Kourkchi}, E., \& {Karachentsev},
  I.~D. 2017, \apj, 835, 78

\bibitem[{{Robitaille} \& {Bressert}(2012)}]{2012ascl.soft08017R}
{Robitaille}, T., \& {Bressert}, E. 2012, {APLpy: Astronomical Plotting Library
  in Python}, Astrophysics Source Code Library, , , ascl:1208.017

\bibitem[{{Sand} {et~al.}(2015){Sand}, {Crnojevi{\'c}}, {Bennet}, {Willman},
  {Hargis}, {Strader}, {Olszewski}, {Tollerud}, {Simon}, {Caldwell},
  {Guhathakurta}, {James}, {Koposov}, {McLeod}, {Morrell}, {Peacock},
  {Salinas}, {Seth}, {Stark}, \& {Toloba}}]{2015ApJ...806...95S}
{Sand}, D.~J., {Crnojevi{\'c}}, D., {Bennet}, P., {et~al.} 2015, \apj, 806, 95

\bibitem[{{Schlafly} \& {Finkbeiner}(2011)}]{2011ApJ...737..103S}
{Schlafly}, E.~F., \& {Finkbeiner}, D.~P. 2011, \apj, 737, 103

\bibitem[{{Shaya} {et~al.}(2017){Shaya}, {Tully}, {Hoffman}, \&
  {Pomar{\`e}de}}]{2017ApJ...850..207S}
{Shaya}, E.~J., {Tully}, R.~B., {Hoffman}, Y., \& {Pomar{\`e}de}, D. 2017,
  \apj, 850, 207

\bibitem[{{Silva} {et~al.}(2005){Silva}, {Massey}, {DeGioia-Eastwood}, \&
  {Henning}}]{2005ApJ...623..148S}
{Silva}, D.~R., {Massey}, P., {DeGioia-Eastwood}, K., \& {Henning}, P.~A. 2005,
  \apj, 623, 148

\bibitem[{{Skrutskie} {et~al.}(2006){Skrutskie}, {Cutri}, {Stiening},
  {Weinberg}, {Schneider}, {Carpenter}, {Beichman}, {Capps}, {Chester},
  {Elias}, {Huchra}, {Liebert}, {Lonsdale}, {Monet}, {Price}, {Seitzer},
  {Jarrett}, {Kirkpatrick}, {Gizis}, {Howard}, {Evans}, {Fowler}, {Fullmer},
  {Hurt}, {Light}, {Kopan}, {Marsh}, {McCallon}, {Tam}, {Van Dyk}, \&
  {Wheelock}}]{2006AJ....131.1163S}
{Skrutskie}, M.~F., {Cutri}, R.~M., {Stiening}, R., {et~al.} 2006, \aj, 131,
  1163

\bibitem[{{Stanimirovi{\'c}} {et~al.}(2006){Stanimirovi{\'c}}, {Putman},
  {Heiles}, {Peek}, {Goldsmith}, {Koo}, {Kr{\v c}o}, {Lee}, {Mock}, {Muller},
  {Pandian}, {Parsons}, {Tang}, \& {Werthimer}}]{2006ApJ...653.1210S}
{Stanimirovi{\'c}}, S., {Putman}, M., {Heiles}, C., {et~al.} 2006, \apj, 653,
  1210

\bibitem[{{Tully} {et~al.}(2016){Tully}, {Courtois}, \&
  {Sorce}}]{2016AJ....152...50T}
{Tully}, R.~B., {Courtois}, H.~M., \& {Sorce}, J.~G. 2016, \aj, 152, 50

\bibitem[{{Tully} \& {Fisher}(1987)}]{1987nga..book.....T}
{Tully}, R.~B., \& {Fisher}, J.~R. 1987, {Nearby galaxies Atlas} (Cambridge:
  University Press, 1987)

\bibitem[{{Tully} {et~al.}(2009){Tully}, {Rizzi}, {Shaya}, {Courtois},
  {Makarov}, \& {Jacobs}}]{2009AJ....138..323T}
{Tully}, R.~B., {Rizzi}, L., {Shaya}, E.~J., {et~al.} 2009, \aj, 138, 323

\bibitem[{{Tully} {et~al.}(2008){Tully}, {Shaya}, {Karachentsev}, {Courtois},
  {Kocevski}, {Rizzi}, \& {Peel}}]{2008ApJ...676..184T}
{Tully}, R.~B., {Shaya}, E.~J., {Karachentsev}, I.~D., {et~al.} 2008, \apj,
  676, 184

\bibitem[{{Vaduvescu} {et~al.}(2005){Vaduvescu}, {McCall}, {Richer}, \&
  {Fingerhut}}]{2005AJ....130.1593V}
{Vaduvescu}, O., {McCall}, M.~L., {Richer}, M.~G., \& {Fingerhut}, R.~L. 2005,
  \aj, 130, 1593

\bibitem[{{Wu} {et~al.}(2014){Wu}, {Tully}, {Rizzi}, {Dolphin}, {Jacobs}, \&
  {Karachentsev}}]{2014AJ....148....7W}
{Wu}, P.-F., {Tully}, R.~B., {Rizzi}, L., {et~al.} 2014, \aj, 148, 7

\end{thebibliography}
\bibliographystyle{aasjournal}
\end{document}